\documentclass[preprint,12pt]{elsarticle}

\usepackage{graphicx}
\usepackage{amsmath,amsfonts,amssymb,lscape,setspace}
\usepackage{amssymb}
\usepackage{lscape}
\usepackage{setspace}
\usepackage{color}
\usepackage{lineno,hyperref,bm,subfigure,amsmath,amssymb}

\journal{}

\begin{document}

\begin{frontmatter}



\title{First-principle study on bulk and (1 1 1) surface of MP (M=K and Rb) in rocksalt structure}


\author[mymainaddress]{Qiang Gao}
\author[mymainaddress]{Lei Li}
\author[mymainaddress]{Huan-Huan Xie}
\author[mymainaddress]{Gang Lei}
 \author[mymainaddress]{Jian-Bo Deng}

\author[mymainaddress]{Xian-Ru Hu\corref{mycorrespondingauthor}}
\cortext[mycorrespondingauthor]{Corresponding author}
\ead{huxianru@lzu.edu.cn}

\address[mymainaddress]{School of Physical Science and Technology, Lanzhou University,
 Lanzhou 730000, People's Republic of China}

\begin{abstract}
The electronic and magnetic properties of bulk and (1 1 1) surfaces for MP (M=K ans Rb) in rocksalt structure have been investigated by employing first-principle calculations. The results reveal that the compounds are half-metallic ferromagnets at the equilibrium lattice constants with large half-metallic band gaps of 0.46 and 0.74 eV. The (111) surfaces of KP and RbP  keep their bulk half-metallic property.  We study the stabilities of the  bulk compounds and their (1 1 1) surfaces as well.

\end{abstract}


\end{frontmatter}


\section{Introduction}
Efficient spin injection from a ferromagnetic metal to a diffusive semiconductor is very meaningful for the development of the performance of spintronic devices \cite{schmidt_fundamental_2000,fert_conditions_2001}. Half-metallic (HM) ferromagnets \cite{katsnelson_half-metallic_2008} are treated as very promising candidates of spin-injector materials for their nearly complete (100\%) spin polarization around the Fermi level. The HM ferromagnet exhibits a metal character in one spin channel and a semiconductor (or insulator） character in the other spin channel, which results in the 100\%  spin polarization around the Fermi level.  In 1983, de Groot first reported HM behavior in Heusler alloy NiMnSb by using first-principle calculations \cite{de_groot_new_1983}. Since then many HM ferromagnets have been predicted theoretically or synthesized experimentally in some kinds of compounds. The full Heusler alloys such as Co$_2$MnAl \cite{rai_study_2010}, half-Heusler alloys such as PtXBi (X=Mn, Fe, Co and Ni) \cite{huang_structural_2015}, quaternary alloys such as CoFeScZ (Z=P, As and Sb) \cite{gao_first-principle_2015}, metallic oxides such as Cr$_2$O \cite{zhang_study_2006} and Fe$_3$O$_4$ \cite{jedema_electrical_2001}, double perovskite compounds such as Pb$_2$FeMoO$_6$ \cite{zhang_half-metallic_2012}, CsCl structure compounds such as CsSe \cite{karaca_half-metallic_2015}, zinc-blende (ZB) structure compounds such as RbX (X = Sb, Te) \cite{senol_half-metallic_2014}, and compounds in rocksalt (RS) structure such as BaC \cite{khalaf_al-zyadi_first-principle_2014} are some of the materials that have been found to be HM ferromagnets by first-principle calculations or some kinds of experiments.

Among the above mentioned HM ferromagnets, contrary to the compounds consist of transition metals, the $p$-electron HM ones have attract much attentions \cite{gao_bulk_2009} for their large HM band gaps guaranteeing the stability of HM property at room temperature. The magnetic moment of the p-electron HM ferromagnet is mainly contributed by the $p$ electrons of the main group elements such as C, N, S and so on, and this is significantly different from the ones with transition metals. The $p$-electron HM ferromagnets have been widely studied. For example MN (M=Ca, Mg, Na, K) \cite{}, MS (M=K, Rb, Cs) \cite{li_first-principles_, rostami_bulk_2013}, and NaX (X=O, S, Se, Te, Po) \cite{ahmadian_half-metallic_2012} have been found to be half-metallicity.

In addition, it is necessary to study the properties of surfaces as the retaining of HM character for surface is very meaningful to the practical applications.  Recent research has shown the half-metallicity of CsCl structure structure RbS and KS is preserved at their (1 1 1) surfaces \cite{li_first-principles_}. As reported in recent paper \cite{khalaf_al-zyadi_first-principle_2014}, the (0 0 1) and (1 1 1) surfaces of RS BaC also retain their bulk HM property. While in the case of ZB CrP, the HM property disappers in the P-terminated (0 0 1) surface and is preserved in the Cr-terminated (0 0 1) surface \cite{lee_first-principles_2006,rahman_magnetism_2007}. So not every surface keeps its bulk HM property. Thus it is very important to investigate the surface properties for the application of HM ferromagnets.

Motivated by the above, we present a first-principle study on the properties of bulk MP (M=K and Rb) in RS structure as well as their (1 1 1) surfaces. It is found that all the bulk compounds are HM ferromagnets at the equilibrium lattice constants. And our calculation results of bulk KP is compared to that have been discussed in Ref \cite{gao_bulk_2009}.  We also investigate the K- and P-terminated (1 1 1) surfaces of KP and Rb- and P-terminated (1 1 1) surfaces of RbP. The results reveal that the bulk half-metallicity are retained in those surfaces.  Finally the stabilities of the bulk compounds and those surfaces are studied.

\section{Methods and details}
\label{sec:1}
The present calculations are performed by using the first-principle full-potential local-orbital (FPLO) minimum-basis method \cite{fplo-1,fplo-2}. The generalized gradient approximation (GGA) is used to treat the exchange-correlation potential \cite{gga-1,gga-2,gga-3}. The scalar relativistic effects are taken into accounts for all calculations. For the Brillouin zone integration, we use the k meshes of
 15$\times$15$\times$15 for bulk RS compounds, and the k meshes of 15$\times$15$\times$1 are adopted for RbP and KP (1 1 1) surfaces. For a self-consistent field iteration, the convergence criterion is set to both the density (10$^{-6}$ in code specific units) and the total energy (10$^{-8}$ hartree). The atomic force convergence criterion is set to 10$^{-3}$\space eV/\AA.

\section{Results and Discussions}
\label{sec:1}
\subsection{The Properties of Bulk MP (M=K, Rb and Cs) in Rocksalt Structrue}
Firstly we calculate the total energy as a function of lattice constants in RS structure for both spin-polarization (FM phase) and non-spin-polarization (NM phase). In \textbf{Fig.1}, we present the curves of total energy versus lattice constant for bulk MP ((M=K and Rb). All the calculations are performed in both FM and NM phases. The calculated results indicate that the FM states are more stable than NM states for the three compounds. The equilibrium lattice constants for KP, RbP and CsP are 6.736 \AA and 7.066 \AA, respectively. The obtained lattice constant for KP is in good agreement with the one in Ref. \cite{gao_bulk_2009}. Therefore the following calculations are performed based on the theoretical equilibrium lattice parameters in RS FM states.

The calculated total density of states (DOS) and partial DOS for MP (M=K and Rb) are shown in Fig.2. As can be seen from the total DOS in Fig.2, the spin-down DOS get through the Fermi level, and it is a metal behaviour. However, there is a large band gap around the Fermi level for each compound in the spin-up band structure, which is an insulator behaviour. So these compounds keep an ideal 100\% spin-polarization of conduction electrons at the Fermi level. And they are HM ferromagnets. From the partial DOS, it shows that the DOS located at the Fermi level are originated exclusively from the 3p orbitals of P. The HM band gaps is defined as the minimum gap between the edge of valence band and Fermi level. So the compounds KP and RbP have HM band gaps of 0.46 eV and 0.74 eV, respectively. The HM band gaps are so large that these compounds may probably keep their HM property at room temperature.

The total magnetic moment is -2.00 $\mu_B$ for each of the three compounds, while the number of valence electrons is 6. So the calculated total magnetic moment per
formula unit, 2.00 $\mu_B$ , for each of the compounds, obeys the Slater-Pauling behaviour which can be expressed by

\begin{equation}
M_{tot}=(Z_{tot}-8) \mu_{B},
\end{equation}

where M$_{tot}$ and Z$_{tot}$ are the total magnetic moment per formula unit and the number of total valence electrons in each compound. To understand the Slater-Pauling behaviour of these compounds, we should investigate the energy levels for the minority-spin  orbitals of them. As can be seen from Fig.2, the spin-up band is the minority-spin state around the Fermi level. So the following discussion is mainly focus on the spin-up band structure. In Fig.3, we present the the fat-bands for bulk KP in RS structure as a representative. The thickness of the fat-bands is proportional to the sum of square of coefficients for these basis states in the expansion of the Kohn-Sham wavefunction into the FPLO orbital basis. The zero energy is responding to the Fermi level. As can be seen from the P-3p and K-3d partial DOS in Fig.2, around the Fermi level the P-3p states have very big wavefunction amplitudes, but the K-3d states almost have no wavefunction amplitudes. From Fig.3, we can see that the P-3s and P-3p (3p-1, 3p+0, 3p+1) states are below the Fermi level. The K-3d (3d-2, 3d-1, 3d+1) state is above the Fermi level. There are no obvious hybridizations between K-3d and P-3p states. It is clear that the energy band gap is created by the maximum of P-3p valence bands and minimum of P-3s valence bands. 

We present one possible schematic representation of the energy levels of the minority-spin band structure for these compounds in Fig.4. As can be seen from Fig.4, there is no orbital hybridization in these compounds. Below the Fermi level, the nondegeneration 3s and triple-degeneration 3p (3p-1, 3p+0, 3p+1)  states are the minority-spin orbitals. So below the Fermi level, the number of minority-spin: $N_{min}=4$
One can directly deduce the number of occupied majority-spin states: $N_{maj}=Z_{tot}-N_{min}$. As the total magnetic moment M$_{tot}$ (in $\mu_B$ ) is just the difference between the number of occupied majority-spin states and occupied minority-spin states. So the the total magnetic moment: $M_{tot}=(N_{maj}-N_{min}) \mu_B=[(Z_{tot}-N_{min})-N_{min}] \mu_B=(Z_{tot}-2N_{min}) \mu_B=(Z_{tot}-8) \mu_B $. Therefore the compounds MP (M=K, Rb and Cs) obey the Slater-Pauling behaviour expressed by Eq.1.

\subsection{The properties to (1 1 1) Surfaces of KP and RbP}
To simulate the (1 1 1) surfaces of KP and RbP, the bulk equilibrium lattice constants of 6.736 \AA \space and 7.066 \AA \space are used. We adopt a slab model containing 15 atomic layers. In order to avoid the interactions of neighboring slabs, a 15 \AA \space vacuum is added above the surfaces.  Each layer includes only one kind of atom, thus there are only two types of terminations, that are named as K (Rb)-terminated
and P-terminated surfaces for KP (RbP). To obtain the equilibrium structure of those slabs, we firstly fix the central layer of the slabs and relax the others and then optimize the slabs by the total energy and atomic force
calculations. After optimization, we realize that the significant change only occurs in the top three layers. And previous study also indicate that the slabs of fifteen layers is adequate to study the properties of the surface in some RS materials \cite{khalaf_al-zyadi_first-principle_2014,gao_surface_2013}. For both the K/P-terminated of KP and the Rb/P-terminated of RbP, the surface atomic layers move toward the center of the slabs, and the second atomic layers move toward the vacuum. However, the positions of third layer atoms change much less than the  surfaces and the second layers, revealing that our model is sufficient for the calculations. Fig.5 and Fig.6 show the calculated surface, subsurface, centre and atomic DOS as well as the total DOS for the relaxed slabs of KP and RbP, respectively. As can be seen from Fig.5 and Fig.6, the canter layer DOS fit quite well with the responding atoms in the bulk compounds. It demonstrates that the physical properties of of bulk compounds are preserved at the centre layers, revealing the models of our slabs is sufficient to study the surface properties. As can be seen from Fig.5 (Fig.6), the HM property of bulk KP (RbP) is retained  by the P- and K-(Rb-) terminated (1 1 1) surfaces as there is a large band gap around the Fermi level in the spin-up band structure while the spin-down DOS get through the Fermi level. The corresponding subsurfaces also keep the bulk HM character. We can directly get that the K(Rb)-terminated (1 1 1) surface shrinks the spin-up band gap of bulk KP (RbP). However the P-terminated surface almost keeps the spin-up band gap of bulk KP (RbP). The two phenomenon is very interesting. We guess that the spin-splitting of K (Rb) atomic layer at the K(Rb)-terminated surface becomes stronger. However, the spin-splitting of P atom is such strong in bulk KP (RbP) that the spin-splitting of P atom at the P-terminated (1 1 1) surface can't change too much.

Table 1 shows the calculated magnetic moments of K and P atoms of KP, Rb and P atoms of RbP, in surfaces and central layers of slabs and also in the bulk for comparisons. In all cases, the atomic magnetic moments of the centre layers fit well with those of the bulk. It means that the  chosen thickness of the slabs is reasonable. This conclusion is consistent with the previous discussion. However, the change of atomic magnetic moment is large for P-terminated (1 1 1) surfaces of both KP and RbP, because the surface atoms lose four of their eight neighbors and K-P, Rb-P bond lengths in these surfaces are changed with respect to those of bulk.

\subsection{Stabilities of Bulk Compounds and KP and RbP Surfaces}
In this section, we will discuss the stabilities of bulk and surfaces for MP (M=K and Rb).

The cohesion energy (E$_{coh}$ ) is a measure of the strength of force that binds atoms together in the solid state which is correlated with the structural
stability in the ground state. The cohesion energy of each alloy per formula unit is expressed

\begin{equation}
E_{coh}=E^M_{atom}+E^P_{atom}-E^{MP}_{total},
\end{equation}

where E$_{tot}$ is the total energy of the considered compound, E$^M_{atom}$ and E$^P_{atom}$ are the the energies of isolated constituent atoms in each compound. The calculated values of cohesion energy are 3.20 eV and 3.05 eV for KP and RbP. The positive cohesion energies indicate the compounds are expected to be stable due to
the positive energies of chemical bonds.

The second aspect of stability for a compound is displayed by the for-
mation energy ($\Delta H_{for}^{MP}$), which shows the stability of the compound with respect
to decomposition into bulk constituents. The formation energy at T=0 K is obtained by considering the reaction to form (or decompose) a crystal from (or into) its components. For each calculated compound, the formation energy of compound can be expressed by

\begin{equation}
\Delta H_{for}^{MP}=E^{MP}_{total}-(E_{M}+E_{P}),
\end{equation}

where E$_{M}$ and E$_{P}$ are the total energies per atom for bulk M (M=K and Rb) and P, respectively. The calculated
values of formation energy are -0.23 eV and -0.14 eV for KP and RbP. In general, a negative value of formation energy indicates the stability of material. All the compounds are likely to be synthesized experimentally due to the negative
formation energy.

We have also considered the structure stability of the two compounds by comparing tetragonal and cubic structures. Fig.7 presents the relationship curves between total energy and c/a  ratio. In our calculations, the cell volumes are fixed at the equilibrium lattice volumes. As can be seen from Fig.7, when c/a=1, both KP and RbP are at the local energy points. So the RS cubic structure is energetically lower than the tetragnoal one. So the  compounds KP and RbP  in RS structure are stable at a certain extent.

In the following, we study the stability of the (1 1 1) surfaces for KP and RbP. The investigation of surface stability is meaningful for the relation of thin films in experiments. The surface energy ($\sigma_{MP}$) is determined as follows:

\begin{equation}
\sigma_{MP}=\frac{1}{A}(
E_{tot}-\mu_{M}^{slab}N_{M}-\mu_{P}^{slab}N_{P}-TS-PV
 ),
\end{equation}
where T, S, P and V are temperature, entropy, pressure, and volume, of which the values are nearly to zero. And they can be ignored. The symbol A is the area of the surface and E$_{tot}$ is the energy of relaxed MP slab. $N_{M}$ and $N_{P}$ are the number of the of the Rb and P atoms in the MP slab, respectively. $\mu_{M}^{slab}$ and $\mu_{M}^{slab}$ are the chemical potentials of the M and P atoms in the slab. The total potentials of M and P atoms in the slab are in equilibrium with that of bulk MP:

\begin{equation}
\mu_{MP}^{bulk}=\mu_{M}^{slab}+\mu_{P}^{slab}.
\end{equation}

Combining Eq.(4) and (5), one can rewrite (4) as follows:

\begin{equation}
\sigma_{MP}=\frac{1}{A}[
E_{tot}-\mu_{M}^{slab}N_{M}-( \mu_{M}^{slab}- \mu_{MP}^{bulk}   )N_{P})],
\end{equation}

The range of chemical potential $\mu_{M}^{slab}$ is limited by

\begin{equation}
\mu_{M}^{bulk}+\Delta H_{for}^{MP}\leq\mu_{M}^{slab}\leq\mu_{M}^{bulk}
\end{equation}

\begin{equation}
\mu_{M}^{slab}\leq\mu_{M}^{bulk}, \mu_{P}^{slab}\leq\mu_{P}^{bulk}
\end{equation}
Here $\Delta H_{for}^{MP}$ has the same meaning as previous. So,

\begin{equation}
\mu_{M}^{bulk}+\Delta H_{for}^{MP}=\mu_{MP}^{bulk}-\mu_{P}^{bulk},
\end{equation}
then

\begin{equation}
\mu_{M}^{bulk}+\Delta H_{for}^{MP}=\mu_{M}^{slab},
\end{equation}

\begin{equation}
\Delta H_{for}^{MP}=\mu_{MP}^{bulk}-\mu_{M}^{bulk}-\mu_{P}^{bulk},
\end{equation}

Combining Eq.(5）and Eq.(8), we can obtain:

\begin{equation}
\Delta H_{for}^{MP}<\mu_{P}^{slab}-\mu_{P}^{bulk}<0.
\end{equation}

The above  expression determines the allowed range of P chemical potential. The formation energy is negative for each compound as has been obtained. Generally, a negative formation energy means that the thin films can be synthesized by experiments in equilibrium conditions. So both the KP and RbP thin films are stable in equilibrium conditions, and can be realized by equilibrium growth techniques. The results of surface energy calculations for KP and RbP (1 1 1) surfaces are shown in Fig.8. The surface energies of all the calculated (1 1 1) slabs have a linear relationship with  P chemical potential. As can be seen from Fig.8, the P-terminated surfaces have positive surface energies in the range of -0.23 to 0 eV and -0.14  to 0 eV for KP and RbP, respectively. However, the surface energies of K- and Rb- terminated are negative in the same ranges. So the K-terminated (1 1 1) surface of KP and Rb-terminated (1 1 1) surface of RbP are much more stable than their P-terminated (1 1 1) surfaces, and may probably be synthesized by experiments.

\section{Conclusions}
\label{sec:2}
In conclusion, we have studied the electronic and magnetic properties of bulk and (1 1 1) surfaces for rocksalt structure MP (M=K and Rb) by first-principle calculations.  It is predicted both the bulk compounds are HM  ferromagnets with large HM band gaps. The Slater-Pauling behaviour of the bulk compounds is investigated in detail as well. The calculations of M (M=K and Rb)-and P-terminated (1 1 1) surfaces reveal that those surfaces keep their  bulk HM character. The magnetic moments of atoms at slabs are compared with those in the responding bulk compound. In addition, the stabilities of the bulk compounds and surfaces are also investigated. Our calculations indicate that the bulk compounds KP and RbP in rocksalt structure are more stable than those in the tetragonal structure. The surface energy calculations show that the K-and Rb-terminated (1 1 1) surfaces are more stable than their P-terminated surfaces. The bulk compounds KP and RbP and their K-and Rb-terminated surfaces may probably be synthesized by experiments for their negative energies. We hope that our work will stimulate experimental efforts to the fabrication of rocksalt structure KP and RbP thin films.





 \bibliographystyle{elsarticle-num}
 \bibliography{Reference.bib}

\newpage
\begin{figure}[!htb]
\centering
\subfigure[]{
\begin{minipage}[c]{0.45\textwidth}
\includegraphics[width=1.6\textwidth,angle=0]{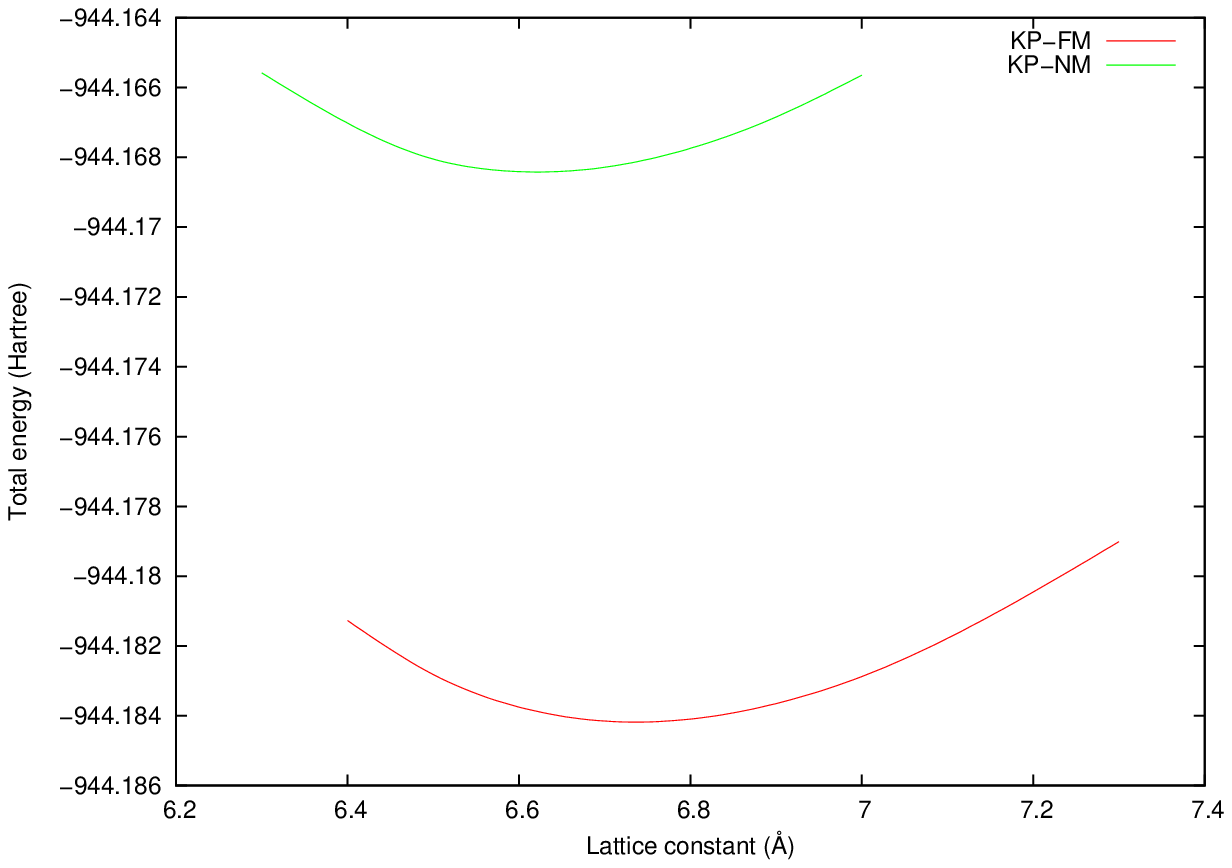} \\
\end{minipage}
}

\subfigure[]{
\begin{minipage}[c]{0.45\textwidth}
\includegraphics[width=1.6\textwidth,angle=0]{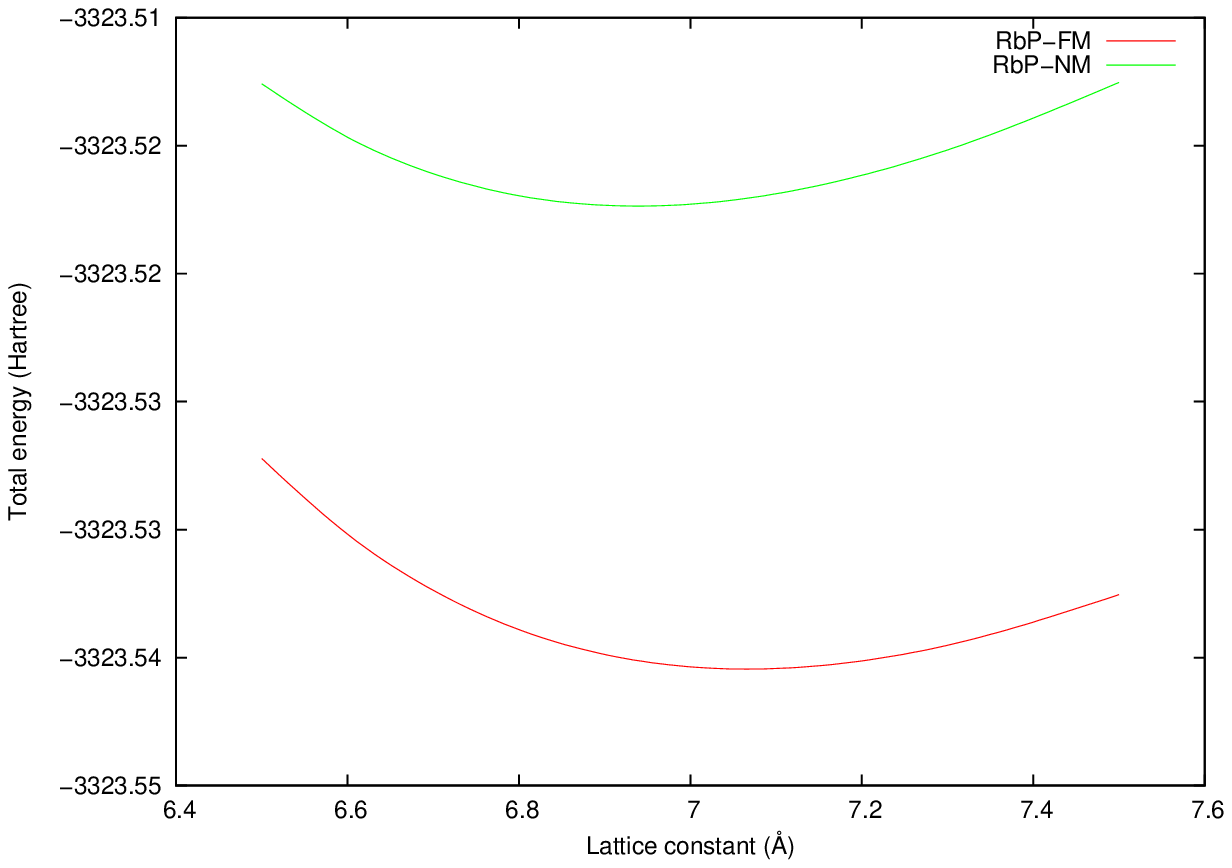} \\
\end{minipage}
}

\caption{}
\end{figure}
\newpage
\begin{figure}[!htb]
\centering
\subfigure[]{
\begin{minipage}[c]{0.45\textwidth}
\includegraphics[width=1.6\textwidth,angle=0]{Figure2-1.eps} \\
\end{minipage}
}

\subfigure[]{
\begin{minipage}[c]{0.45\textwidth}
\includegraphics[width=1.6\textwidth,angle=0]{Figure2-2.eps} \\
\end{minipage}
}\caption{}
\end{figure}
\newpage
\begin{figure}[!htb]
\centering
\subfigure[]{
\begin{minipage}[c]{0.45\textwidth}
\includegraphics[width=1.6\textwidth,angle=-90]{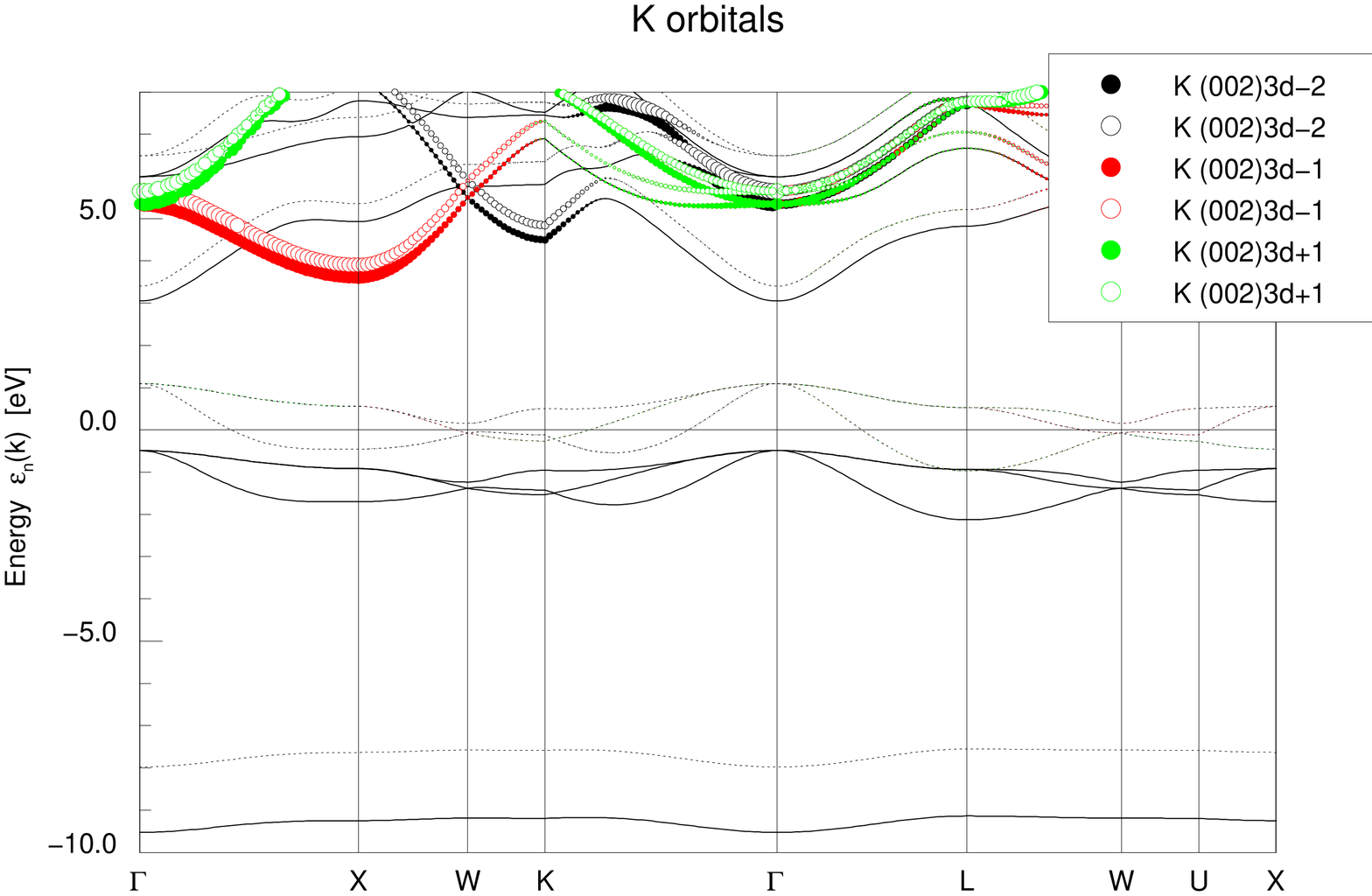} \\
\end{minipage}
}

\subfigure[]{
\begin{minipage}[c]{0.45\textwidth}
\includegraphics[width=1.6\textwidth,angle=-90]{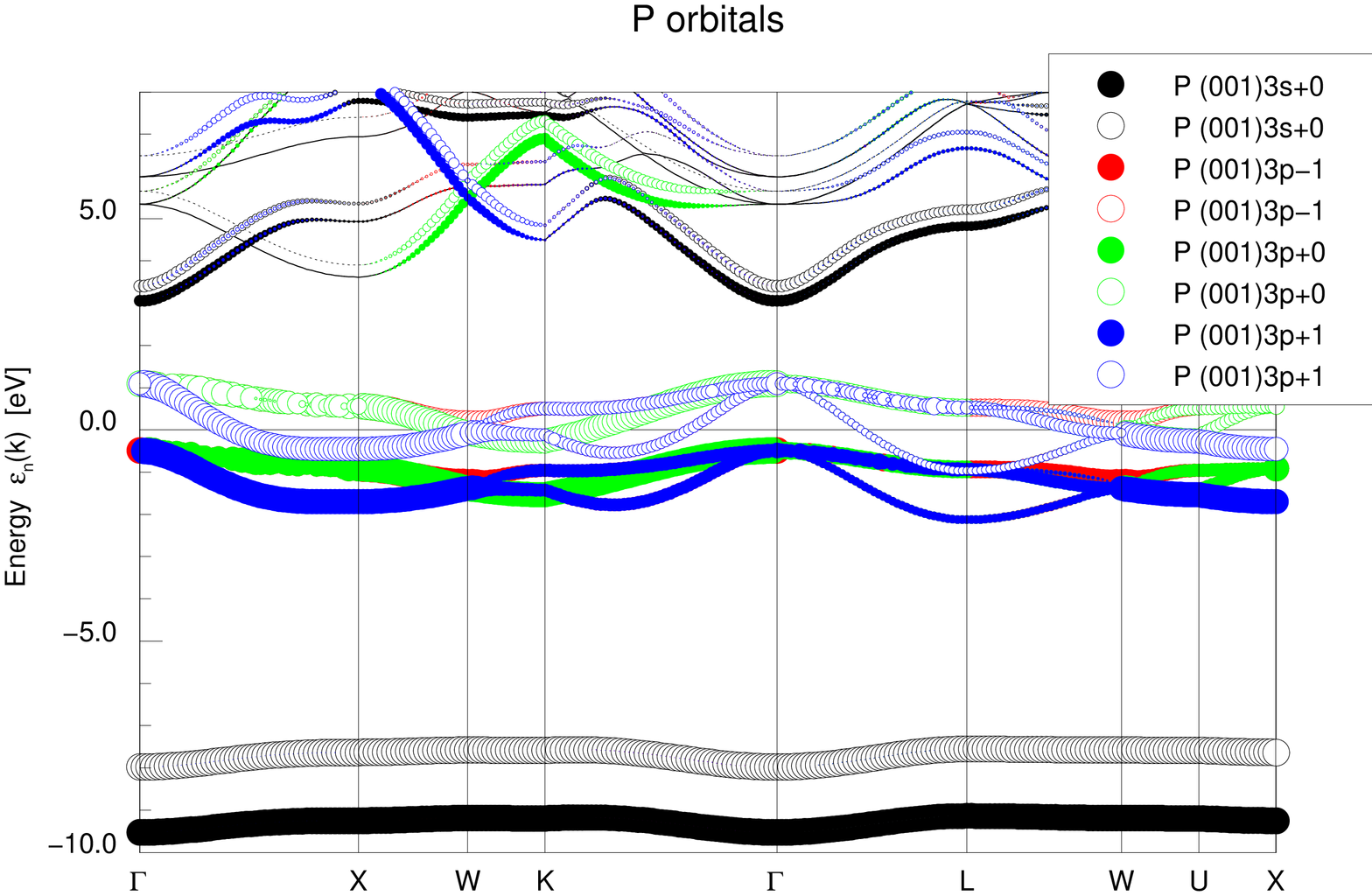} \\
\end{minipage}
}

\caption{}
\end{figure}
\newpage
\begin{figure}[htp]
\centering
\includegraphics[scale=0.60]{Figure4.eps}
\caption{}
\label{}
\end{figure}
\newpage
\begin{figure}[htp]
\centering
\includegraphics[scale=0.70]{Figure5.eps}
\caption{}
\label{}
\end{figure}
\newpage
\begin{figure}[htp]
\centering
\includegraphics[scale=0.70]{Figure6.eps}
\caption{}
\label{}
\end{figure}
\newpage
\begin{figure}[!htb]
\centering
\subfigure[]{
\begin{minipage}[c]{0.45\textwidth}
\includegraphics[width=1.6\textwidth,angle=0]{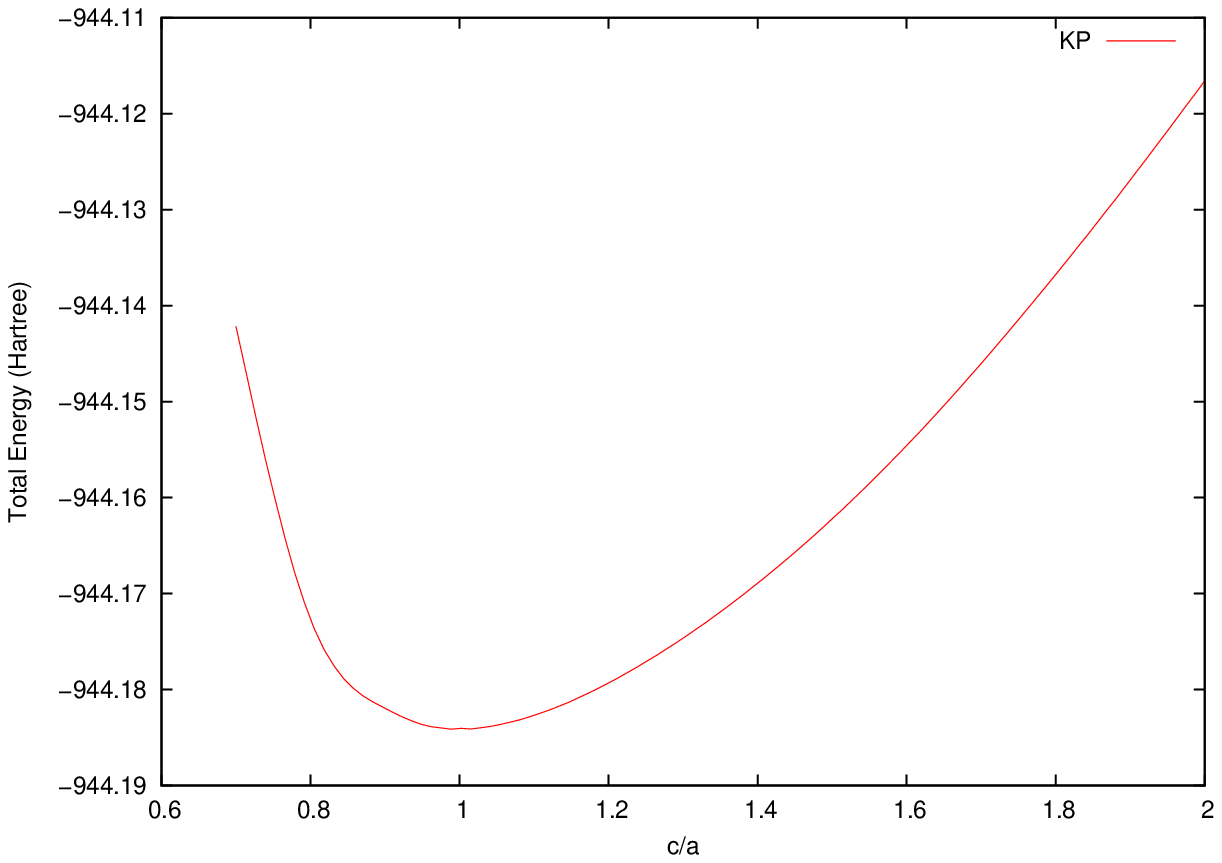} \\
\end{minipage}
}

\subfigure[]{
\begin{minipage}[c]{0.45\textwidth}
\includegraphics[width=1.6\textwidth,angle=0]{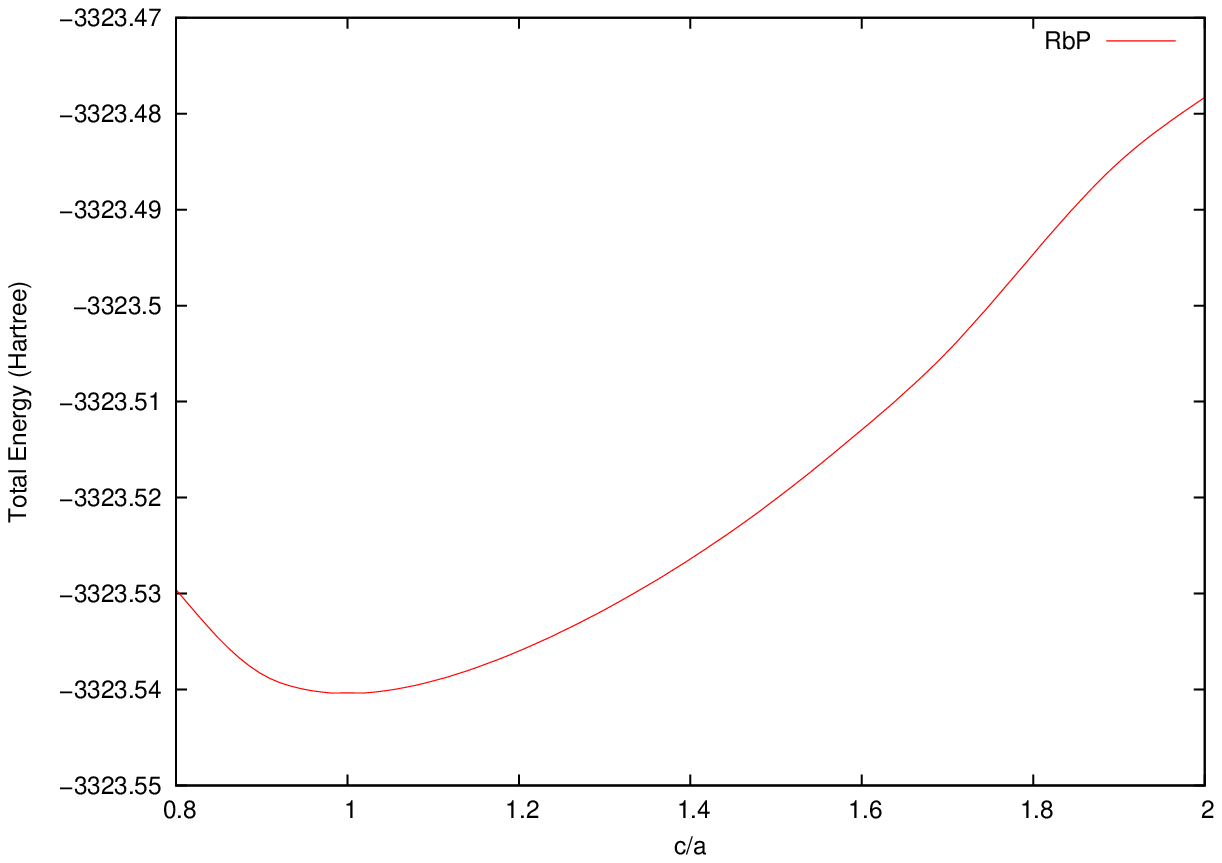} \\
\end{minipage}
}
\caption{}
\end{figure}
\newpage

\begin{figure}[!htb]
\centering
\subfigure[]{
\begin{minipage}[c]{0.45\textwidth}
\includegraphics[width=1.6\textwidth,angle=0]{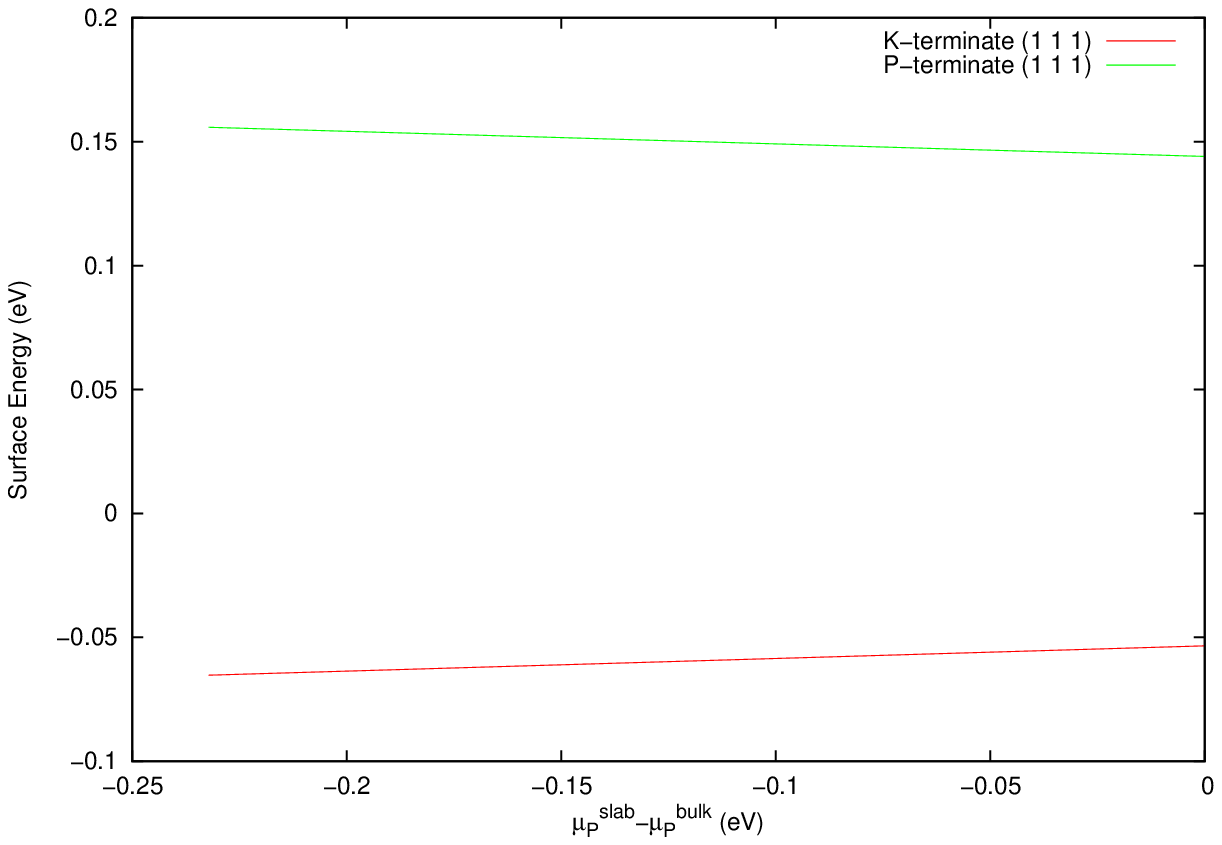} \\
\end{minipage}
}

\subfigure[]{
\begin{minipage}[c]{0.45\textwidth}
\includegraphics[width=1.6\textwidth,angle=0]{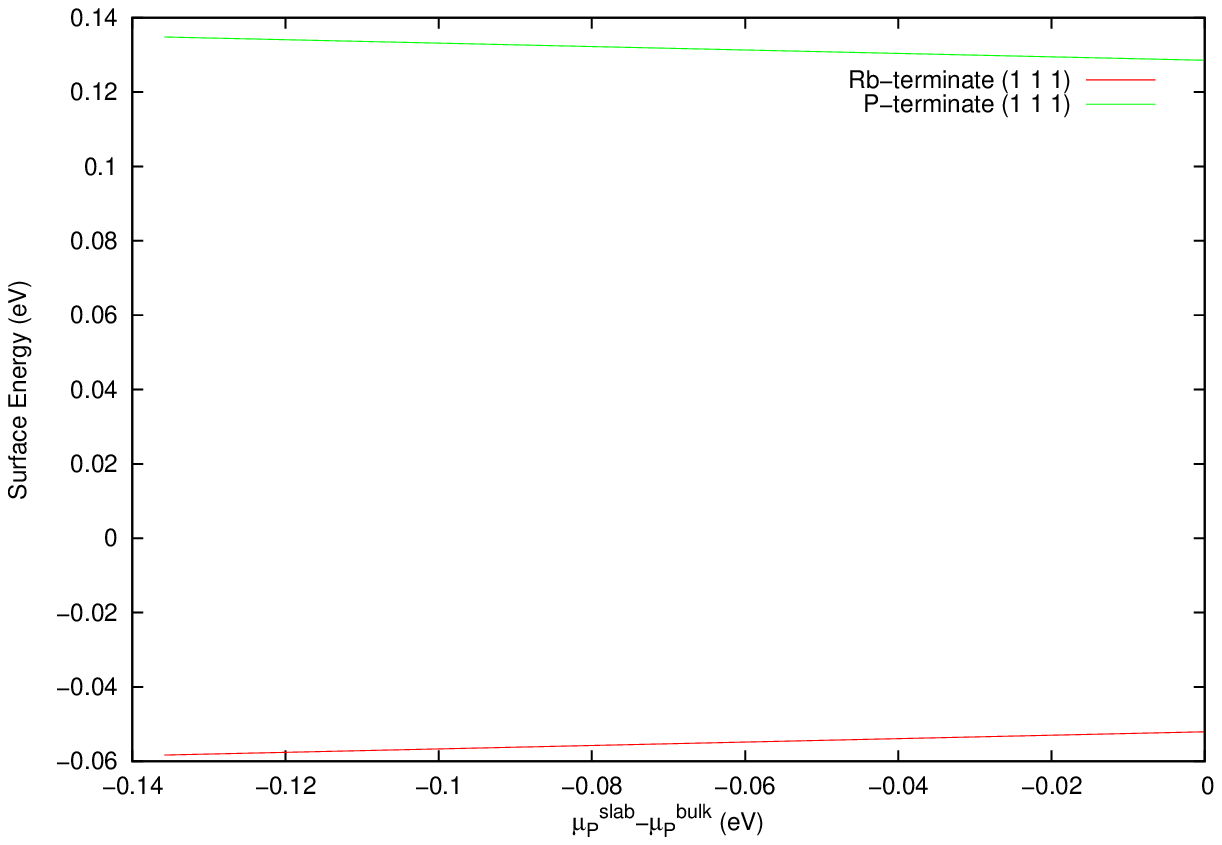} \\
\end{minipage}
}
\caption{}
\end{figure}

\begin{table}[ht]
\newcommand{\tabincell}[2]{\begin{tabular}{@{}#1@{}}#2\end{tabular}}
\caption{The magnetic moments of Rb and S atoms, K and S atoms in different positions of the (111) slabs and in the bulk of RbS, KS.} 
\centering 
\begin{tabular}{c c c c c c c c c c c} 
\hline\hline 
MP & \tabincell{c}{Atomic \\species} & \tabincell{c}{position of \\the atom} & \tabincell{c}{M-(111)\\terminated} & \tabincell{c}{P-(111)\\terminated} &  \\ [0.5ex] 
\hline 
KP & \tabincell{c}{K \\K\\K\\P\\P\\P} & \tabincell{c}{Surface \\Centre layer\\Bulk\\Surface \\Centre layer\\Bulk} & \tabincell{c}{-0.051 \\- \\ -0.065 \\ - \\2.054  \\ 2.065 } &\tabincell{c}{- \\-0.052 \\ -0.065 \\2.556  \\  -\\  2.065} \\ 
\hline 
RbP & \tabincell{c}{Rb \\Rb\\Rb\\P\\P\\P} & \tabincell{c}{Surface \\Centre layer\\Bulk\\Surface \\Centre layer\\Bulk} & \tabincell{c}{-0.065  \\-  \\-0.063  \\-  \\2.055  \\2.063  } &\tabincell{c}{- \\-0.054 \\ -0.063 \\2.560  \\ - \\2.063  } \\ [1ex]
\hline \hline
\end{tabular}
\label{table:nonlin} 
\end{table}

\newpage

\newpage
\textbf{Figures}

\end{document}